# Gauge invariant investigation of the nature of Confinement

A. González-Arroyo and A. Montero [a]*

[a]Dpto. de Física Teórica C-XI, Univ. Autónoma de Madrid,
Cantoblanco, Madrid 28049, SPAIN

We observe a strong correlation between the decrease in the number of action density peaks in SU(2) Yang-Mills configurations with cooling and that of the string tension. The nature and distribution of these peaks is investigated. The relationship with monopole currents after the abelian projection is also considered.

## 1. Introduction

In this paper we report on some of our more interesting results within our program to analyse the physical content of the SU(2) Yang-Mills vacuum. Our final goal is to achieve a comprehensive description of the structure of the vacuum and, hence, an *understanding* of its more salient properties: Confinement, chiral symmetry breaking, etc. A similar goal is pursued by other groups with contributions collected in these Proceedings. Our work, nonetheless, focuses on gauge invariant structures and is hence different, but not, necessarily antagonistic, from attempts based on the abelian projection.

Our methodology is, in broad terms, standard. We have Monte Carlo generated 56 configurations on an $8^3 \times 64$ lattice with twisted boundary conditions in space ( $\vec{m} = (1,1,1)$ ) and Wilson action with $\beta = 2.325$. Then, we have cooled these configurations using over-improved cooling [1] and $\epsilon = -0.3$. Finally, we have studied the gauge-invariant properties of the resulting configurations. Before proceeding to describe our results, let us clarify some technical points about our method.

The spatial size of our lattice in physical units is $l_s = 1.25\,fm.$ ( with $\sigma = 5\,fm^{-2}$ ). Although at this size there are probably non-negligible finite-size effects, we claim that qualitatively the structure of the vacuum is essentially the same as for infinite volume. We base this belief on the results of our previous M.C. simulations at this same point [2], where we measured the string tension $\sigma$

*Work financed by CICYT grant AEN93-0693

from correlations of Polyakov loops with different windings in space. We got results consistent with the infinite volume string tension with up to 10% differences and errors. Both the $l_s$ dependence and the equality of the estimates from correlators with different windings (electric fluxes) are in agreement with the string formation picture.

Another point of concern is the effect of cooling. Contrary to other authors, we do not try to argue that the results obtained are unaffected by this technique. Rather, we have carried an independent study to understand the effect that cooling might have on the results. We have looked at previously prepared configurations, cooled them for a while, and observed the outcome. Since cooling is a local minimization technique it is clear that classical solutions will sooner or later dominate the configurations. Thus, we first examined the effect on individual instantons. Contrary to the case of the continuum, instantons on the lattice are not invariant under cooling. The degeneracy of zero-modes is broken by artifacts and cooling produces a motion along this valley. In the standard cooling algorithm, instantons become small and eventually disappear through the holes of the lattice. Our proposal of over-improved cooling makes use of the fact that one can change the size and the sign of artifacts to make this effect smaller and eliminate the instability of sufficiently large instantons. In Ref. [3] we advocated that this variant cooling method could be used to obtain a better determination of the topological susceptibility. (One poster in this Conference [4] presents a more refined proposal within this idea.) In this work we have chosen $\epsilon = -0.3$. With this choice, large instantons are stable, but at the same time motion along the zero-mode valley is not too fast. Concerning individual instantons, hence, we conclude that our cooling essentially simply removes those which are smaller than 2 lattice spacings. In Fig. 1 we show the topological susceptibility as a function of the number of cooling steps for $\epsilon = -0.3$ and for naive cooling ($\epsilon = 1$).

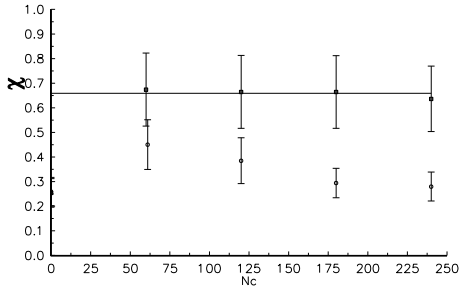

Figure 1. The topological susceptibility as a function of cooling step for $\epsilon = -0.3$ (closed squares) and ordinary cooling (white circles).

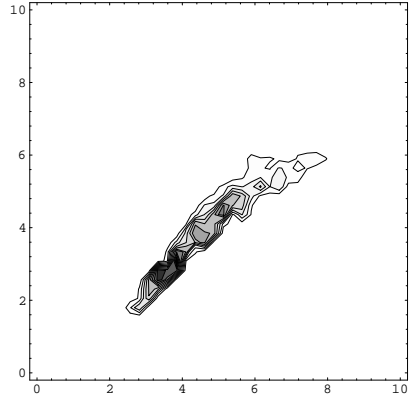

Figure 2. Distribution of the peaks in the $\rho_1$ (vertical) and $\rho_2$ (horizontal) plane.

We have also investigated the effect of cooling on instanton-antiinstanton pairs. We have observed that cooling produces the annihilation of such a pair, although the rate at which this happens depends widely on separation and sizes of the instantons. This mechanism is independent on the value of $\epsilon$ and, hence, cannot be eliminated. We should then understand its effect and correct for it.

Now we should consider which are going to be our main observables. We will look for *peaks* (i.e. local maxima) in the action density, its electric and magnetic parts separately and the topological charge density. For each peak we will record as attributes, its location, its height and its width, obtained from the first neighbours of the peak position. We aim at an understanding of the nature of these peaks and a description of their distribution from our data. Some of these peaks will be instantons, of course. Since they depend on a single scale $\rho$, for them the height and the width are related. On the continuum we can form two independent functions of height and width $\rho_1$ and $\rho_2$ as follows:

$$\rho_1 = \left(\frac{48}{S(0)}\right)^{\frac{1}{4}} \qquad (1)$$

$$\rho_2 = \sqrt{\frac{-32 S(0)}{\triangle S(0)}}, \qquad (2)$$

where S(x) is the action density for the configuration with a peak located at $x = 0$ and $\triangle S(0)$ its 4-dimensional laplacian at this point. For a BPST instanton $\rho_1$ and $\rho_2$ give the same value and equal to the instanton size parameter $\rho$.

## 2. Results

In a previous publication [3], we showed the behaviour of the string tension and the topological susceptibility as a function of the number of cooling steps. As reported by many authors before, both quantities decrease under the naive cooling method. As mentioned previously, if we use overimproved cooling and $\epsilon = -0.3$ the decrease in the topological susceptibility is eliminated (see Fig.

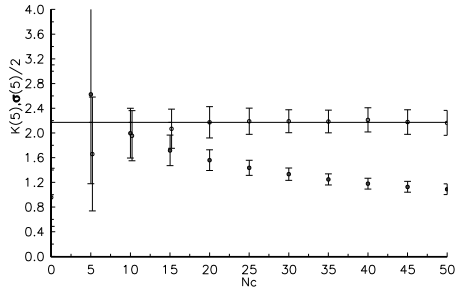

Figure 3. The quantity $K(5) =$ (open circles) versus the number of cooling steps. The effective string tension $\frac{\sigma(5)}{2}$ in $fm^{-2}$ units (closed circles) is also shown.

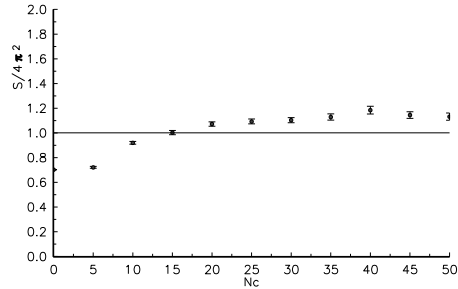

Figure 4. We plot the average action of a configuration per peak divided by $4\pi^2$ for each cooling step.

1), but however that is not the case for the string tension, which at 50 cooling steps has dropped to 40% of its uncooled value. Let us clarify what is meant by *string tension* for the cooled configurations. We may measure an effective string tension $\sigma(t)$ through correlations of Polyakov loops at distances $t$ and $t-1$. This quantity shows first a plateau for large but finite $t$, which is the quantity we are referring to. Nonetheless, if $t$ is large enough ($t > N_c$), as argued by Teper, we should recover the uncooled string tension.

In the present work, as mentioned previously, we studied the nature and distribution of peaks. The first result concerns the degree of self-duality observed. The peaks in the electric action density coincide in 90% of the cases with those in the magnetic density (or the topological charge density). The coincidence is in number, location and height. Usually, if spurious or non-symmetric peaks appear, they seem to be associated with transient processes of instanton-antiinstanton annihilation. Hence, from now on we will not specify which peaks we are referring to, since they all give the same answers. Our next observation, is that the number of peaks decreases with the number of cooling steps $N_c$. From 25 to 50 coolings this number has dropped by a factor of 2. It is quite natural to interpret this decrease as due to our previously mentioned instanton-antiinstanton annihilation mechanism. But, are our peaks BPST instantons? To see this, we checked the equality of $\rho_1$ and $\rho_2$ for our peaks. In Fig 2. the contour plot in the $\rho_2 \rho_1$ plane is given. Most peaks are lying along the diagonal, with deviations due to finite-lattice corrections. This is a non trivial fact, which need not be the case even for self-dual peaks. We conclude that our peaks are locally instanton-like. They need not be BPST instantons, since there are other configurations showing the same behaviour. In particular, the $Q = \frac{1}{2}$ instanton of Ref. [5] is known to be present and to behave in this way.

An important observation that we made, is the existence of a complete correlation between the decrease in the density $D$ of peaks and that of the string tension. We may form a dimensionless quantity $K = \frac{\sigma}{\sqrt{D}}$ and study its evolution with the number of cooling steps. The result is invariant under cooling to a 2% level from 20 to 50 cooling steps. Being more precise, we may write $K(t)$ for the quantity obtained from the effective mass $\sigma(t)$. Our statement applies for all values of $t$ from 2 to 6. In Fig. 3 we show $K(5)$.

The value of $\frac{\sigma(5)}{2}$ in $fm^{-2}$ units is also plotted, to show how the string tension decreases in this same interval. If we fit a constant value number to the values obtained from 20 to 50 cooling steps, we get $K(5) = 2.18 \pm 0.205 \pm 0.027$ where the first error is statistical (for a single cooling step), and the second is the maximum difference observed in absolute value between the constant and the value for all cooling steps from 20 to 50. For other $t$ we get: $K(2) = 0.939 \pm 0.024 \pm 0.027$, $K(3) = 1.463 \pm 0.048 \pm 0.037$, $K(4) = 1.881 \pm 0.097 \pm 0.036$, $K(6) = 2.331 \pm 0.427 \pm 0.028$, $K(7) = 2.236 \pm 0.813 \pm 0.048$. Notice that for the larger values of $t$, $K(t)$ tends to a constant (which is what we call string tension).

In Ref. [6,2] we proposed an scenario for the Yang-Mills vacuum based on a gas of self-dual $|Q| = \frac{1}{2}$ lumps with mean separation $\bar{d}$ around $0.7 fm$. This picture predicts precisely that the string tension should scale with the square root of the density $\bar{d}^{-2}$. Using $\sigma = 5 fm^{-2}$ and $\bar{d} = 0.7\ fm$, we get $K = 2.5$, which is close to the estimates from our data. Our model has obviously a crucial test: the observed lumps should be associated with $\pm \frac{1}{2}$ topological charge. To check this point, we computed the total action of each of our configurations and divided it by the number of peaks of that configuration. The value obtained was in all cases close to $4\pi^2$ (the value for a self-dual $|Q| = \frac{1}{2}$ object). Averaging over all configurations, we obtain the mean action per peak as a function of cooling steps, which is plotted in Fig. 4. Notice the small errors, reflecting the small dispersion of the data. The value is also fairly stable under cooling given the large variation of numerator and denominator in the 20 to 50 step region. We lack a simple algorithm that would tell us what peak is associated to $Q = \frac{1}{2}$ and which to a normal instanton. Nonetheless we possess a method which is non-local and quite lengthy to be used, but can be used for a few configurations. These control configurations at 100-200 cooling steps are very clearly seen to be made of a majority of $|Q| = \frac{1}{2}$ lumps and a few integer charge ones. For fewer number of steps the identification is less beautiful and requires taking into account transient instanton-antiinstanton annihilation states as well.

We conclude with a summary and comments. Our data provides clear evidence for the correlation between the string tension, as measured through Polyakov loops, and the density of peaks. The latter decreases with cooling via pair annihilation (also $|Q| = \frac{1}{2}$ pairs annihilate), and drives the former. The observed correlation is to be expected within our picture of the vacuum based on a gas of $|Q| = \frac{1}{2}$ lumps. Some evidence for the presence of these lumps has been given, both in an statistical fashion as by monitoring individual configurations. A more detailed description of our results can be found in Ref [7]. Let us mention to conclude, that we have looked (together with M. Polikarpov) at what results from the $|Q| = \frac{1}{2}$ configurations after the abelian projection. When using Maximal abelian gauge, we found a single monopole line going through the center of the lump. This is again the case if one takes extended monopoles of size 2a. For other gauge fixings (Polyakov lines or plaquette) a more involved structure of monopole lines is found. More work in this direction is in progress. We conclude that, presumably, a gauge-invariant description of Confinement is not in contradiction with the Dual Superconductor picture.